\documentclass{article}
\usepackage{graphicx} 
\usepackage[a4paper,left=2.5cm,right=2.5cm,top=3cm,bottom=3cm]{geometry}
\usepackage{amsthm}
\usepackage{amsmath}
\usepackage{amssymb}
\usepackage{hyperref}
\usepackage{graphicx}
\usepackage{tabularx}
\usepackage{tikz}
\usepackage{caption}
\usepackage{comment}
\usepackage{blindtext}
\usepackage{setspace}
\usepackage[authoryear]{natbib}
\usepackage{pgfplots}
\pgfplotsset{compat=1.18}
\usepackage{authblk}

\newtheorem{proposition}{Proposition}   

\title{Optimal vs. Naive Diversification in the Cryptocurrencies Market: The Role of Time-Varying Moments and Transaction Costs}
\author{ Heming Chen$^{\dagger}$, Xiaojing Cai$^{\ddagger}$}
\date{October, 2025}

\begin{document}

\maketitle

\begingroup
\renewcommand\thefootnote{}
\footnotetext{$^{\dagger}$ Faculty of Economics, Okayama University, Tsushimanaka 3-1-1, Kita-ku, Okayama 700-8530, Japan. E-mail: pet26ubl@s.okayama-u.ac.jp}
\footnotetext{$^{\ddagger}$ Corresponding author: Faculty of Economics, Okayama University, Tsushimanaka 3-1-1, Kita-ku, Okayama 700-8530, Japan.
Graduate School of Economics, Kobe University, 2-1, Rokkodai, Nada-Ku, Kobe 657-8501, Japan. E-mail: caicai@okayama-u.ac.jp}
\endgroup

\begin{abstract}
This study investigates three central questions in portfolio optimization. First, whether time-varying moment estimators outperform conventional sample estimators in practical portfolio construction. Second, whether incorporating a turnover penalty into the optimization objective can improve out-of-sample performance. Third, what type of optimal portfolio strategies can consistently outperform the naive $1/N$ benchmark. Using empirical evidence from the cryptocurrencies market, this paper provides comprehensive answers to these questions. In the process, several additional findings are uncovered, offering further insights into the dynamics of portfolio construction in highly volatile asset classes.
\end{abstract}

\section{Introduction}
\subsection{Related literature}
Markowitz’s modern portfolio theory was proposed in 1952. Since then, it has attracted extensive attention from both academia and the financial industry. However, in many practical applications, the Markowitz rule and its variants even perform worse than equally weighted naive diversification. \citet{michaud1989markowitz} explains this tendency using the “error-maximizing” property of mean–variance optimization, which suggests that the error in estimating risk and return leads to poor performance. Under the assumption that excess return follows a multivariate normal distribution, by using an expected loss function in the standard mean–variance analysis context, \citet{kan2007optimal} analytically show that the classical plug-in method using sample estimates to replace true parameters in optimization problems can result in poor out-of-sample performance. Under the same assumption, \citet{demiguel2009optimal} indicate that a minimum length of estimation window is required such that a standard mean–variance strategy can outperform the $1/N$ strategy because of the existence of parameter uncertainty or estimation error. 
\par
As estimation error is one of the main causes of poor performance generated by optimal diversification, how to address estimation error has become a significant issue. In fact, there has been extensive literature devoted to this issue in academia. For instance, \citet{demiguel2009optimal} evaluate the performance of a standard mean–variance model and its 13 extensions, covering almost all prominent models proposed in previous studies to mitigate the impact of estimation error. Most of these extensions employ the Bayesian or various shrinkage approaches, and the rest impose certain restrictions on the estimated moments. Nevertheless, according to the empirical results, \citet{demiguel2009optimal} report that none of the considered models can consistently outperform naive diversification for seven empirical datasets, which raises serious questions about the usefulness of portfolio theory. Fortunately, by using a shrinkage estimator approaching $1/N$, \citet{tu2011markowitz} develop a combination rule that optimally combines the $1/N$ weights with the weights obtained from each of four sophisticated rules derived from investment theory and find that all of them outperform both the original sophisticated rules and the naive $1/N$ rule. This result seems to vindicate the mean–variance theory. Furthermore, \citet{kirby2012s} revisit the results of \citet{demiguel2009optimal} and find that the poor performance is largely because of their research design, which places mean–variance optimization at an inherent disadvantage. By setting the conditional expected return equal to that of the $1/N$ rule, the mean–variance model can outperform the $1/N$ rule for most of the datasets used by \citet{demiguel2009optimal} when transaction costs are not considered. In the presence of transaction costs, \citet{kirby2012s} also propose two new timing strategies dominating naive diversification, which mitigates estimation error by solely exploiting the estimated conditional volatility or estimated reward-to-risk ratios. 
\par
Most of the aforementioned studies focus primarily on improving the misspecification of the first two moments of asset returns through statistical analysis under the assumption of normality, while \citet{kirby2012s} effectively implement a new perspective on time-varying moments. There has been ample empirical evidence indicating the existence of time-vary asset return moments \citep{gao2018commodities}. A considerable number of asset allocation studies adopt a time-varying perspective, with some studies focusing on volatility timing (see, e.g., \citet{fleming2001economic} and \citet{fleming2003economic}), others emphasizing return forecasting (see, e.g., \citet{ahmed2016can}, \citet{opie2020global}) , and most concerned with both matters (see, e.g., \citet{della2009economic}, \citet{kirby2012s} and \citet{gao2018commodities}). In recent years, there has been a growing number of asset allocation studies employing machine-learning techniques for return forecasting (see, e.g. \citet{d2020artificial}, \citet{chen2021mean}, \citet{ma2021portfolio}, \citet{kynigakis2022does} and \citet{du2022mean}).
\par

However, in the real world, any gains from optimal diversification can be easily eroded by large transaction costs. In traditional mean–variance optimization problems, the gains obtained from a rebalancing may not compensate the costs required for a rebalancing. Thus, it becomes necessary to introduce transaction costs into the optimization problem. \citet{yoshimoto1996mean} propose an optimization system with a V-shaped cost function, and the empirical results show that ignoring transaction costs results in higher turnover and an inefficient portfolio. \citet{olivares2018robust} theoretically and empirically explore the role of transaction costs in portfolio optimization problems. They prove that the mean–variance problem with p-norm transaction costs is equivalent to three different problems designed to deal with the estimation error: a robust portfolio problem, a regularized linear regression problem, and a Bayesian portfolio problem. The data-driven approach proposed by \citet{olivares2018robust} also typically outperforms the benchmark portfolio because it addresses transaction costs and estimation error simultaneously. Similarly, \citet{hautsch2019large} first theoretically demonstrate the regulatory effect of quadratic and proportional transaction costs. Then an extensive empirical study in a large-scale portfolio optimization framework shows that the ex ante incorporation of transaction costs is crucial for achieving reasonable portfolio performance. 
\par

Building on the aforementioned studies, our analysis is organized around two key comparisons within the portfolio optimization framework. The first comparison investigates whether conditional moment estimators based on a time-varying perspective can achieve superior performance relative to conventional sample estimators. The second comparison examines whether incorporating a turnover penalty term into the objective function enhances portfolio performance compared with the standard mean–variance formulation. Specifically, for the conditional mean estimation, we employ probabilistic time-series forecasting based on a state-of-the-art deep-learning method. For the covariance estimation, we employ a dynamic conditional correlation (DCC)-EGARCH model to capture the volatility dynamics. The turnover penalty term is further introduced to account for transaction costs and to mitigate excessive portfolio rebalancing.
\par

In particular, we consider a special case in which all mean estimates are set to zero, corresponding to the global minimum-variance portfolio. \citet{demiguel2009optimal} find that the minimum-variance strategy successfully reduces the extreme weights and the turnover of the portfolio relative to the mean–variance strategy. \citet{olivares2018robust} also observe that the minimum-variance strategy generally outperforms the mean–variance strategy, which they explain by the difficulties in estimating mean returns. Following numerous studies employing daily data, \citet{hautsch2019large} also ignore the estimation of mean returns but perform the minimum-variance strategy. 
\par

Consequently, we consider four types of portfolio optimization objective functions, depending on whether a turnover penalty is included and whether the mean estimate is incorporated. Furthermore, for each specification, the optimization can be implemented using either time-varying moment estimators or sample estimators, allowing for a comprehensive comparison across both modeling and estimation dimensions. 
\par

We employ cryptocurrencies as the empirical dataset in this study to evaluate the performance of the aforementioned portfolio strategies. As an emerging asset class, cryptocurrencies have attracted growing attention from investors, regulators, and academics alike. \citet{platanakis2018optimal} conduct an empirical investigation using weekly data on four major cryptocurrencies and conclude that there is no significant difference between the naive $1/N$ rule and the sample-based mean–variance strategy in the cryptocurrencies market. 
\par

\citet{katsiampa2019high} employ a diagonal BEKK model to capture the dynamic conditional covariance structure and show that time-varying conditional correlations exist. In addition, a growing body of literature provides evidence that cryptocurrency returns are also predictable (e.g., \citep{liu2021risks}, \citet{kraaijeveld2020predictive}, and \citet{fousekis2021directional}). These findings underscore the necessity of accounting for time-varying moments in the construction of cryptocurrency portfolios. Another important feature of cryptocurrencies is their extremely high volatility, roughly an order of magnitude greater than that of traditional financial assets (\citep{bariviera2021we}), which in turn highlights the importance of considering transaction costs. This consideration motivates us to focus on the cryptocurrencies market as the empirical testing ground.
\par

To the best of our knowledge, this is the first study that systematically investigates how time-varying moment estimators and turnover penalty jointly affect portfolio performance in the cryptocurrencies market. In addition, we employ data with different time frequencies and various levels of risk aversion to enhance the robustness of our findings across alternative investment horizons and investor preferences. Through a comprehensive empirical analysis, this study provides novel evidence on the relative importance of estimation dynamics, transaction costs, and temporal span in portfolio construction.
\par

\subsection{Contribution and preview}

This study contributes to the literature on portfolio optimization primarily in the following ways.
\par
First, we emphasize the impact of the variation levels of different estimators, which capture the temporal changes of a series between consecutive periods, rather than its dispersion around the mean as measured by variance. Most previous studies focus solely on the forecasting accuracy of estimators (see, e.g., \citet{d2020artificial}, \citet{chen2021mean}, \citet{ma2021portfolio} and \citet{du2022mean}). However, \citet{fleming2001economic} show that volatility timing works better with smoother covariance estimates than with those obtained from the minimum MSE criterion. \citet{fleming2003economic} observe that volatile multivariate GARCH estimates result in inferior performance of volatility timing. \citet{kirby2012s} consider an alternative estimator of conditional means with a lower asymptotic variance, which in turn reduces portfolio variance and turnover. This study focuses more on the transaction costs driven by mutable characteristics than on portfolio variance, so we assess different moment estimators by their variation levels rather than variances.
\par

Second, given numerous theoretical and empirical benefits of the turnover penalty \citep{yoshimoto1996mean,olivares2018robust,hautsch2019large}, we incorporate a turnover penalty into portfolio construction. Furthermore, we theoretically examine the role of rebalancing frequency in shaping the impact of the turnover penalty on portfolio performance. Despite the differences in analytical frameworks, our finding is consistent with the deduction in \citet{woodside2013portfolio} (the expected portfolio return per period is given by the weighted sum of asset returns, minus the transaction cost scaled by investment horizon H), which suggests that the impact of introducing transaction costs diminishes as H increases.
\par

Third, we adopt the larger one of two analytical solutions for performance fee as our evaluation criteria in accordance with \citet{kirby2012s} and elaborate on its economic implication to justify this selection. We emphasize this issue because directly applying some software packages may yield the smaller analytical solution, which is economically unreasonable. We explicitly show that higher return and lower risk relative to the benchmark lead to a positive and larger performance fee. Moreover, return plays a more significant role than risk in determining the performance fee.
\par

Fourth, rather than focusing solely on the final evaluation metrics, we examine in detail the impact of different alternative models on several key aspects that determine portfolio performance. Most prior research has mainly concentrated on the final performance metrics (see, e.g., \citet{demiguel2009optimal}, \citet{della2009economic}, \citet{gao2018commodities}, \citet{ahmed2016can} and \citet{opie2020global}). On a risk-adjusted basis, portfolio performance depends on return and risk, with the net return decomposed into gross return and transaction costs. We therefore analyze how the alternative models in this study affect the three key determinants of portfolio performance: gross return, transaction costs, and risk.
\par

To preview our results:

(\MakeUppercase{\romannumeral 1}) In portfolio optimization, the time-varying estimators of the first- and second-moment (mean and covariance) are both inferior to their sample counterparts, regardless of whether a turnover penalty is imposed. As time-varying estimators (deep-learning forecasts and DCC covariances) are typically more volatile, they incur substantial transaction costs while failing to improve asset allocation or minimize portfolio risk relative to their sample counterparts. Our finding is consistent with and extends the discussions in \citet{fleming2001economic}, \citet{fleming2003economic}, \citet{kirby2012s}, \citet{gao2018commodities} and \citet{kynigakis2022does}.
\par

(\MakeUppercase{\romannumeral 2}) Regarding the turnover penalty, our empirical results show that it improves transaction costs for all optimal strategies but fails to improve the final performance of every strategy. The exception arises in the volatility-timing strategy employing sample covariances, where the turnover penalty deteriorates portfolio performance. This outcome closely aligns with the results reported for the short-sale-constrained minimum-variance portfolio with nominal transaction costs in \citet{olivares2018robust}. However, imposing the turnover penalty leads to a notable improvement in volatility-timing 
 portfolios employing DCC covariances, enabling them to achieve performance comparable to those using sample estimates, thereby extending the results of \citet{fleming2001economic} and \citet{fleming2003economic}. Within the mean–variance framework, our empirical results highlight the importance of the turnover penalty in reducing transaction costs and enhancing performance, in line with \citet{yoshimoto1996mean} and \citet{olivares2018robust}. Furthermore, our analytical derivation suggests that the improvement resulting from the turnover penalty diminishes as the rebalancing frequency decreases, which is consistent with \citet{woodside2013portfolio}. This pattern is also confirmed by our empirical results.
\par

(\MakeUppercase{\romannumeral 3}) Even in the highly correlated cryptocurrencies market, most volatility-timing portfolios achieve positive performance fees and outperform the naive $1/N$ benchmark, underscoring the economic value of the predictability of asset return volatility for portfolio construction. Even after the turnover penalty is imposed, portfolios utilizing return estimates remain inferior to both the volatility-timing portfolios and the naive $1/N$ benchmark, reconfirming that it is difficult to exploit the potential predictability of asset returns to generate investment gains. These results are consistent with and extend many previous studies, such as \citet{olivares2018robust} and \citet{hautsch2019large}
\par

The remainder of this paper is organized as follows. Section 2 introduces the four optimization frameworks employed in this study and analytically derives the role of rebalancing frequency. Section 3 presents the methodology for estimating the conditional moments of asset returns and describes the dataset used. Section 4 details the optimization modeling process and the various performance measures. Section 5 reports and discusses the empirical results for both daily and weekly rebalancing cases. Finally, Section 6 concludes the paper.

\section{Optimization frameworks}
In this section, we present four optimization frameworks for determining optimal portfolio weights and discuss the regularization effects introduced by the turnover penalty.

\subsection{MV and MVC optimization}
Building on the aforementioned studies, we adopt a time-varying framework for portfolio optimization. As a starting point, we introduce a conditional mean–variance optimization approach (hereafter referred to as ``MV'' optimization), similar to that in \citet{della2009economic} and \citet{kirby2012s}:
\begin{equation}
\begin{aligned}
& \underset{w_t}{min} \; \{ \frac{\gamma}{2}w_t'\Sigma_{t+1|t}w_t - w_t'\mu_{t+1|t} \} \\
& s.t. \; w_t'l=1 
\end{aligned}
\end{equation}
where $w_t$ is the N-dimensional decision variable representing the optimal portfolio weights estimated at time t, $\mu_{t+1|t}$ and $\Sigma_{t+1|t}$ are the conditional mean and conditional covariance matrix of N risky asset returns from time t to time t+1 given the information set $I_{t}$, respectively, $l$ is a $N \times 1$ vector of ones, and $\gamma$ denotes the coefficient of relative risk aversion. In this study, we evaluate portfolio performance using risk-aversion coefficients ranging from 1 to 10 for all optimal strategies, where “strategy” refers to a specific combination of mean estimator, covariance estimator, and optimization framework, or the naive 1/N rule.
\par
Given the numerous theoretical and empirical benefits of the turnover penalty, we consider the second optimization framework incorporating transaction costs into the objective function (``MVC'' optimization, hereafter). \citet{olivares2018robust} argue that the quadratic transaction costs ($L_2$-norm) are more suitable to handling estimation error than the proportional transaction costs ($L_1$-norm), although the latter is more realistic. However, they also point out that if there is no estimation error, then proportional transaction costs are actually optimal. In this study, we aim to examine whether our alternative approaches improve the forecasting accuracy of conditional moments, and hence we consider the MVC optimization framework with proportional transaction costs as follows:
\begin{equation}
\begin{aligned}
& \underset{w_t}{min} \; \{ \frac{\gamma}{2}w_t'\Sigma_{t+1|t}w_t + \beta||w_t - w_{t-1^{+}}^*||_{1} - w_t'\mu_{t+1|t} \} \\
& s.t. \; w_t'l=1
\end{aligned}
\end{equation}
where $ \beta = 0.005 $ represents proportional transaction costs of 50 basis points for each of the risky assets \citep{demiguel2009optimal}, $w_{t-1^{+}}^*$ denotes the vector of portfolio weights before rebalancing at time t (with the initial value set to $\vec 0$), and $\beta||w_t - w_{t-1^{+}}^*||_{1}$ represents the transaction costs incurred during rebalancing at time t. It is noteworthy that asset prices have changed from time t to time t+1, and thus $w_{t-1^{+}}^*$ differs from the weights by which the portfolio was rebalanced at time t-1 (i.e. optimal portfolio weights derived by the optimization framework at time t-1, namely $w_{t-1}^*$). Formally, $w_{t-1^{+}}^* = (w_{1,t-1^{+}}^*,\cdots,w_{i,t-1^{+}}^*,\cdots,w_{N,t-1^{+}}^*)$ with $w_{i,t-1^{+}}^*$ representing the portfolio weight of asset i before rebalancing at time t, which can be expressed as follows:
\begin{align}
w_{i,t-1^{+}}^* = \frac{(1+r_{i,t})'w_{i,t-1}^*}{\sum_{i=1}^{N}(1+r_{i,t})'w_{i,t-1}^*}, \ \forall i.
\end{align}
where $r_{i,t}$ is the return of asset i during the period from time t-1 to time t, and $w_{i,t-1^{+}}^*$ is defined as above. In the MVC optimization framework, transaction costs are subtracted from the objective function, and hence potential penalties are automatically taken into account when rebalancing the portfolio. Proposition \ref{prop1} shows that the regularization effect resulting from the turnover penalty varies with the rebalancing frequency.
\begin{proposition}
\label{prop1}
The MVC optimization framework is equivalent to the classical mean–variance optimization problem with: 
\begin{equation}
\begin{aligned}
& \underset{w_t}{min} \; \{ \frac{\gamma}{2}w_t'\Sigma_{t+1|t}w_t - w_t'\tilde{\mu}_{t|t+1} \} \\
& s.t. \; w_t'l=1
\end{aligned}
\end{equation}
where $\tilde{\mu}_{t|t+1}=\mu_{t+1|t}-\beta g^*$, and $g^*$ is the subgradient vector of function $|| w_{t} - w_{t-1^+}^* ||_1 $ evaluated at $w_t^*$. 
\end{proposition}
As shown in Proposition \ref{prop1}, the turnover penalty in the MVC optimization framework implies a regularization effect that shifts the conditional mean by $\beta g^*$. As $g^*$ is bounded by $\pm1$, the lower the rebalancing frequency, the smaller $\beta g^*$ will be relative to $\mu_{t+1|t}$, and hence the smaller the change caused by the turnover penalty becomes. Intuitively, this also makes sense, because the lower the rebalancing frequency, the smaller the influence of transaction costs, which is consistent with \citet{woodside2013portfolio}.

\subsection{GMV and GMVC optimization}
As the impact of estimation error is largely due to errors in mean estimation \citep{demiguel2009optimal}, we consider two optimization frameworks that rely solely on the conditional covariance estimates. The first is the global minimal variance portfolio (``GMV'' optimization, hereafter).
\begin{equation}
\begin{aligned}
& \underset{w_t}{min} \; \{\frac{\gamma}{2}w_t'\Sigma_{t+1|t}w_t  \}  \\
& s.t. \; w_t'l=1 
\end{aligned}
\end{equation}
Due to volatile nature of the volatility estimates, we also consider the influence of the turnover penalty and propose the variance-cost optimization framework (``GMVC'' optimization, hereafter) for portfolio selection, which is similar to the approach of \citet{olivares2018robust}.
\begin{equation}
\begin{aligned}
& \underset{w_t}{min} \; \{\frac{\gamma}{2}w_t'\Sigma_{t+1|t}w_t + \beta||w_t - w_{t-1^{+}}^*||_1 \}  \\
& s.t. \; w_t'l=1 
\end{aligned}
\end{equation}
The GMVC optimization framework minimizes the sum of risk (volatility) and transaction costs without utilizing estimation of the conditional mean, and thus is not influenced by the associated estimation error. \citet{hautsch2019large} demonstrate that it is equivalent to the traditional GMV optimization framework with the objective function $w'\Sigma_{\frac{\beta}{\gamma}}w$, where the $\Sigma_{\frac{\beta}{\gamma}}$ is defined as follows:
\begin{align}
\Sigma_{\frac{\beta}{\gamma}}=\Sigma + \frac{\beta}{\gamma}(g^*l'+lg^{*'}).
\end{align}
The turnover penalty in the GMV optimization framework implies a regularization effect that shifts conditional covariance by $\frac{\beta}{\gamma}(g^*l'+lg^{*'})$. Similar to the previous analysis, the lower the rebalancing frequency, the smaller $\frac{\beta}{\gamma}(g^*l'+lg^{*'})$ will be relative to $\Sigma$ and hence the smaller the change induced by the turnover penalty.
\par

Due to estimation error, mean–variance analysis often leads to extreme weights that are far from optimal \citep{demiguel2009optimal}. Hence, we impose the following short selling constraint across all four optimization frameworks mentioned above in the empirical analysis presented below:
\begin{align}
w_i \geq 0, \forall i
\end{align}
In addition to the constraint $w_t'l=1$, we restrict the weights to lie between 0 and 1.

\section{Data and estimation approaches}
In this section, we begin by introducing the dataset employed in our study and the data pre-processing procedures. Subsequently, we describe three approaches for mean estimation and two approaches for volatility estimation.

\subsection{Data and pre-processing}
The dataset employed in this study comprises daily and weekly returns for the four longstanding and most liquid cryptocurrencies over the entire observation period: Bitcoin, Ethereum, Ripple and Litecoin, obtained from \url{http://www.coinmarketcap.com}. We first collected daily closing price data in US dollars over the period from 7th Aug 2015 to 14th July 2023, as this start date corresponds to the earliest point at which price data for all four cryptocurrencies are available. Then the log return in cryptocurrency i during the period from time t to time t+1 can be written as:
\begin{align}
    r_{i,t+1} = ln(p_{i,t+1}) - ln(p_{i,t} ), \ \forall i.
\end{align}
where $p_{i,t+1}$ and $p_{i,t}$ are closing prices of cryptocurrency i at time t and t+1 respectively (for daily data $p_{t+1}$ is the closing price of the next day, while for weekly data $p_{t+1}$ is the closing price seven days later). The samples consist of 2898 observations for daily returns and 414 for weekly returns, for each of the four cryptocurrencies. We then define wealth immediately prior to portfolio rebalancing at time t+1 as $W_{t+1}$, which can be calculated as:
\begin{align}
W_{t+1} = W_{t}(1-\beta||w_t^* - w_{t-1^+}^*||_1)(1+r_{t+1}'w_t^*),
\end{align}
where $W_t$ is similarly defined as wealth immediately prior to portfolio rebalancing at time t, $\beta||w_t^* - w_{t-1^+}^*||_1$ represents the transaction costs incurred during portfolio rebalancing at time t, and $r_{t+1} = (r_{1,t+1},\ldots,r_{N,t+1})'$ is the vector of returns for the cryptocurrencies considered. The aforementioned $\mu_{t+1|t}$ and $\Sigma_{t+1|t}$ are the conditional mean and conditional covariance matrices of $r_{t+1}$, respectively. Subsequently, the portfolio return net of transaction costs used for performance evaluation during the period from time t to time t+1 (defined as $R_{p,t+1}^{*}$) is calculated as:
\begin{equation}
\begin{aligned}
R_{p,t+1}^{*} = \frac{W_{t+1}}{W_{t}} - 1 &= (1-\beta||w_t^* - w_{t-1^+}^*||_1)(1+r_{t+1}'w_t^*) - 1 \\ 
&\approx r_{t+1}'w_t^* - \beta||w_t^* - w_{t-1^+}^*||_1.
\end{aligned}    
\end{equation}
After omitting higher-order terms, the portfolio return net of transaction costs is approximately equal to the gross portfolio return ($r_{t+1}'w_t^*$) minus transaction costs ($\beta||w_t^* - w_{t-1^+}^*||_1$). Accordingly, transaction costs are explicitly incorporated in performance evaluation.

\subsection{Sample mean and covariance matrix}
While $w_t$ is the decision variable in the optimization framework, $\mu_{t+1|t}$ and $\Sigma_{t+1|t}$ need to be input into the optimization problem as given parameters. However, the true values of these two parameters are usually unknown and hence need to be estimated. Traditionally, the sample mean and sample covariance matrix are used as their estimates, defined as follows:
\begin{align}
\hat{\mu}_{t+1|t} &= \frac{1}{M}\sum_{j=0}^{M-1}{r_{t-j}}, \\
\hat{\Sigma}_{t+1|t} &= \frac{1}{M}\sum_{j=0}^{M-1}{(r_{t-j}-\hat{\mu}_{t+1|t})(r_{t-j}-\hat{\mu}_{t+1|t})' }.
\end{align}
where M is the length of the estimation window.

\subsection{Multivariate GARCH}
In addition to the sample estimators discussed above, there are also other methods to forecast the conditional covariance matrix, typically including various multivariate GARCH models. For the conditional mean equation, we follow \citet{katsiampa2019high} and adopt a random walk model with drift, specified as follows:
\begin{align}
    r_t = \mu + \varepsilon_t,
\end{align}
where $r_t$ is defined as above, $\mu$ is a constant vector representing the mean returns of the assets, and $\varepsilon_t$ is a vector of residuals with a conditional covariance matrix $\Sigma_t$. \citet{cheikh2020asymmetric} document an inverse asymmetry effect in the variances of major cryptocurrencies, which means that positive shocks tend to increase volatility more than negative shocks of the same magnitude, contrary to what is typically observed in traditional financial assets. We thus employ a DCC model \citep{engle2002dynamic} to model multivariate volatility, with the exponential GARCH or EGARCH \citep{nelson1991conditional} used for a univariate GARCH estimation process to capture the inverted asymmetric effect as follows:
\begin{align}
\varepsilon_{i,t} &= \sqrt{h_{ii,t}}v_{i,t}, \\
ln(h_{ii,t}) &= \omega + \alpha_1(\frac{\varepsilon_{t-1}}{h_{ii,t-1}^{0.5}}) + \alpha_2|\frac{\varepsilon_{t-1}}{h_{ii,t-1}^{0.5}}| + \beta_1ln(h_{ii,t-1}),
\end{align}
where $h_{ii,t}$ is the conditional variance of cryptocurrency i (the diagonal element of $\Sigma_t$), and $v_{i,t}$ is the corresponding standardized residual. $ln(h_t)$ responds asymmetrically to positive and negative shocks because of the presence of the term $\varepsilon_{t-1}/h_{t-1}^{0.5}$. In addition, neither the sign of $ln(h_t)$ nor that of the estimated parameters is restricted. Thus, this model can capture not only the asymmetry effect but also the inverse asymmetry effect. It is not difficult to verify that the covariance matrix $\Sigma_t$ can be expressed as:
\begin{align}
    \Sigma_t = D_t R_t D_t.
\end{align}
where $R_t = [\rho_{ij,t}]$ is the conditional correlation matrix with $\rho_{ij,t}=h_{ij,t}/(h_{ii,t}h_{jj,t})^{0.5}$, and $D_t = diag(h_{11,t}^{0.5}, \ldots,h_{NN,t}^{0.5})$ is a diagonal matrix where $h_{ii,t}$ can be estimated from the last step (denoted by $\hat{h}_{ii,t}$). The conditional correlation matrix $R_t$ is then estimated by using a smoothing process. Finally, the estimated model is employed to generate one-step-ahead forecasts, yielding the estimate of $\Sigma_{t+1|t}$. 

\subsection{Deep-learning approaches}
In portfolio optimization problems, the impact of errors in estimating means is significantly greater than that of errors in estimating variances and covariances, and variances-covariances can be estimated more accurately than means when the number of assets is not too large \citep{chopra1993effect,ackermann2017optimal}. Therefore, we put more effort into estimating the conditional mean (expected return) and employ the two best-performing individual deep-learning approaches in \citet{makridakis2023statistical}: DeepAR and SimpleFeedForward, implemented in the GluonTS toolkit to perform one-step-ahead return forecasts. GluonTS is a Python package for probabilistic time-series modeling based on deep learning, with many state-of-the-art models built in. Originally proposed by \citet{salinas2020deepar}, DeepAR is a probabilistic forecasting method based on autoregressive recurrent neural networks (RNNs). SimpleFeedForward is a simple and fast multi-layer perceptron (MLP) model, however it often outperforms more complex architectures, as shown in \citet{makridakis2023statistical}. It is worth noting that the implementations in GluonTS may differ slightly from the original models due to certain adaptations.
\par

Probabilistic models return a representation of a probability distribution rather than simple point forecasts, and we can extract any needed statistic from sample paths representing the probability distribution \citep{alexandrov2019gluonts}. Probabilistic forecasting surpasses traditional point forecasting in two key aspects: (1) it is more appropriate for capturing the inherent randomness of many time series; (2) it provides a measure of a model's predictive uncertainty \citep{li2024deepar}. \citet{golnari2024probabilistic} propose a deep-learning model based on probabilistic gated recurrent units (P-GRU) for cryptocurrency price forecasting, and find that probabilistic forecasting outperforms traditional approaches. In addition, the probabilistic models in GluonTS are trained using a ``cross-learning'' approach, which means that the model is trained using all available time series rather than on each individually, thereby exploiting the advantage of estimating parameters globally (\citep{januschowski2020criteria, makridakis2023statistical}).

\section{Modeling process}
In this section, we begin by presenting the descriptive statistics of our dataset, to justify the validity of our statistical modeling framework. We then outline the procedures for hyper-parameter tuning and deep-learning forecasting. Subsequently, we summarize the complete portfolio selection pipeline. Finally, we describe the performance evaluation metrics employed in this study, with particular emphasis on the performance fee.

\begin{table}[htbp]
    \refstepcounter{table}
    \renewcommand{\arraystretch}{1} 
    \caption*{Table 1: Descriptive statistics of cryptocurrency returns employed in this study.}
    \phantomsection\label{Table 1}
    \begin{tabularx}{\textwidth}{cXXXXXXX}
        \hline
        \multicolumn{8}{l}{Panel A: Results for daily returns} \\
        \hline
        & Min & Max & Growth(\%)\footnotemark[1] & Mean & Std Dev & Skewness & Kurtosis \\
        \hline
        BTC & -0.4647 & 0.2251 & 10749.87 & 0.001617 & 0.03816 & -0.7479 & 11.27535 \\
        ETH & -1.3029 & 0.4104 & 69859.60 & 0.002260 & 0.06230 & -3.1540 & 71.57900 \\
        XRP & -0.6164 & 1.0275 & 8722.37 & 0.001546 & 0.06546 & 
        2.1735 & 36.44553 \\
        LTC & -0.4490 & 0.5114 & 2161.25 & 0.001076 & 0.05389 &
        0.2879 & 11.37623\\
        \hline
        \multicolumn{8}{l}{Panel B: Results for weekly returns} \\
        \hline
        & Min & Max & Growth(\%)\footnotemark[1] & Mean & Std Dev & Skewness & Kurtosis \\
        \hline
        BTC & -0.4945 & 0.4119 & 10749.87 & 0.01132 & 0.1011 & -0.2578 & 2.443796 \\
        ETH & -0.6034 & 0.8849 & 69859.60 & 0.01582 & 0.1564 & 0.8351 & 4.892879 \\
        XRP & -0.6080 & 1.0985 & 8722.37 & 0.01082 & 0.1717 & 1.7542 & 7.479678 \\
        LTC & -0.5935 & 0.8759 & 2161.25 & 0.00753 & 0.1394 & 0.6952 & 5.664237 \\
        \hline
    \end{tabularx}
\end{table}
\footnotetext[1]{The ``Growth(\%)'' represents the price growth of individual cryptocurrencies over the entire observation period.}

\subsection{Descriptive statistics and statistical modeling}

The descriptive statistics of the asset returns used in this study are shown in \hyperref[Table 1]{Table 1}. The mean returns of all cryptocurrencies are positive for both weekly or daily data. For both frequencies, Ethereum has the highest mean return (0.23\% and 1.58\%), while Litecoin has the lowest mean return (0.11\% and 0.75\%). Similarly, for both weekly and daily data, Ripple has the highest standard deviation (6.55\% and 17.17\%), while Bitcoin has the lowest standard deviation (3.82\% and 10.11\%). Significant skewness and kurtosis are observed for the returns of both frequencies except for the kurtosis of weekly BTC returns (2.44) being less than 3, and the kurtosis of weekly returns is notably lower than that of daily returns. 
\par
Moreover, the results of the Jarque-Bera test in \hyperref[Table 2]{Table 2}, which reports several statistical tests on the characteristics of the return series, also reject the normality hypothesis for all return series. For daily returns, both ARCH(8)-PQ and ARCH(8)-LM tests show strong evidence supporting the presence of ARCH effects. For weekly returns, although the ARCH(8)-PQ test results are not significant, the ARCH(8)-LM test results strongly confirm the presence of ARCH effects. Accordingly, it is appropriate to model the volatility dynamics using the multivariate GARCH framework described in Section 2, assuming a multivariate Student’s \textit{t} distribution.
\par
Following \citet{demiguel2009optimal}, we employ the rolling-sample approach for statistical modeling. Specifically, regardless of daily or weekly returns, the first 70\% of the dataset is used as the training set to fit the multivariate GARCH model and the remaining 30\% is used as the testing set for out-of-sample forecasting and evaluation. Let T denote the length of the total return series and Q denote the out-of-sample length. For each out-of-sample time t, we use the previous T-Q returns to estimate the parameters of multivariate GARCH models, and one-step-ahead forecasts are then obtained from the fitted model. Similarly, the sample estimators are obtained in the same manner while the length of estimation window M is no longer set to 70\% of the dataset. Following \citet{platanakis2018optimal}, we set M = 26 for the weekly rebalancing case, and for the daily case, we set M = 182 (=26$\times$7).
\par

We then compute the unconditional correlation matrices of cryptocurrency returns using the testing set data. As shown in \hyperref[Table 3]{Table 3}, all correlation coefficients are greater than 0.5, implying a highly correlated market. The pursuit of an optimal strategy that can outperform the naive $1/N$ strategy in such a highly correlated market constitutes the main motivation of this study.
\par

\begin{table}[htbp]
    \refstepcounter{table}
    \renewcommand{\arraystretch}{1} 
    \caption*{Table 2: Statistical tests.}
    \phantomsection\label{Table 2}
    \begin{tabularx}{\textwidth}{cXXXX}
        \hline
        \multicolumn{5}{l}{Panel A: Results for daily returns} \\
        \hline
            & Phillips-Perron test & Jarque-Bera test & ARCH(8)-PQ & ARCH(8)-LM \\
        \hline
        BTC & -55.184*** & 15649*** & 96.16003*** & 3190.133*** \\
         & (0.01) & ($<$ 2.2e-16) & (0.000000e+00) & (0)\\
        ETH & -57.127*** & 624369*** & 54.62043*** & 1776.4245*** \\
         & (0.01) & ($<$ 2.2e-16) & (5.231947e-09) & (0) \\
        XRP & -55.637*** & 162913*** & 329.2337*** & 2928.2261*** \\
         & (0.01) & ($<$ 2.2e-16) & (0) & (0) \\
        LTC & -54.674*** & 15695*** & 150.2582*** & 2901.4734*** \\
         & (0.01) & ($<$ 2.2e-16) & (0) & (0) \\
        \hline
        \multicolumn{5}{l}{Panel B: Results for weekly returns} \\
        \hline
            & Phillips-Perron test & Jarque-Bera test & ARCH(8)-PQ & ARCH(8)-LM \\
        \hline
        BTC & -18.86*** & 109.88*** & 14.229457* & 153.57064*** \\
         & (0.01) & ($<$ 2.2e-16) & (0.0759773) & (0.000000e+00) \\
        ETH & -18.722*** & 467.92*** & 14.13467* & 236.66240*** \\
         & (0.01) & ($<$ 2.2e-16) & (0.07832218) & (0.000000e+00) \\
        XRP & -15.609*** & 1192.1*** & 80.81058*** & 159.59806*** \\
         & (0.01) & ($<$ 2.2e-16) & (3.352874e-14) & (0.000000e+00) \\
        LTC & -19.785*** & 595.27*** & 6.856076 & 292.67549*** \\
         & (0.01) & ($<$ 2.2e-16) & (0.55223778) & (0.000000e+00) \\
        \hline
    \end{tabularx}
    \begin{tabular}{p{\textwidth}}
    Note: Values in parentheses are the p-values. *,**, and *** indicate significance at the 10\%, 5\% and 1\% levels, respectively.
    \end{tabular}
\end{table}

\subsection{Hyper-parameter selection and deep-learning forecasting}

In this subsection, we determine suitable hyper-parameters for our dataset and forecasting task. Due to computational resource constraints, we focus on several of the most important hyper-parameters for optimization. Following \citet{makridakis2023statistical}, we adopt a set of indicative values for the selection process. Specifically, because the output of probabilistic deep-learning models is probability distributions, we take the average of ``num\_samples'' sample paths as the point prediction input into the optimization problem, where the hyper-parameters ``num\_samples'' denote the number of samples used to construct the model. The hyper-parameter ``context\_length'' which determines the number of time steps considered for computing predictions, is set to 26 for weekly returns and 182 for daily returns to ensure consistency with the sample estimators. \footnotemark[2] We employ the Optuna library in Python to conduct hyper-parameter tuning via a (tree-structured Parzen estimator) TPE algorithm \citep{bergstra2011algorithms}. 
\footnotetext[2]{The explanations of the hyper-parameters are obtained from \url{https://ts.gluon.ai/stable/index.html}.}
\par

We use the same rolling-sample and dataset splitting approaches as in statistical modeling. However, the initial 70\% of the dataset is divided into two parts at this stage, the first 60\% serves as the training set for model training, and the remaining 10\% is used as the validation set for hyper-parameter selection. After selecting the optimal hyper-parameters, the first 70\% of the dataset is used again as the final training set, while the remaining 30\% is reserved for model performance evaluation. We employ aggregate root-mean-square error (RMSE) as the optimization objective aggregated both across time steps and time series, which can be expressed as:
\begin{align}
aggregate \ RMSE = \sqrt{\frac{\sum_{t=1}^{L}\sum_{i=1}^{N}{(r_{i,t}-\hat{r}_{i,t})^2}}{NL}}.
\end{align}
where L denotes the length of the validation set and $\hat{r}_{i,t}$ denotes the deep-learning forecast for cryptocurrency. Additional details on the hyper-parameter selection are provided in Appendix B. 

\begin{table}[htbp]
    \refstepcounter{table}
    \renewcommand{\arraystretch}{1} 
    \caption*{Table 3: Unconditional correlation matrices.}
    \phantomsection\label{Table 3}
    \begin{tabularx}{\textwidth}{c|XXXX|XXXX}
        \hline
        & \multicolumn{4}{l|}{Panel A: Results for daily returns} & \multicolumn{4}{l}{Panel B: Results for weekly returns} \\
        \cline{2-9}
            & BTC & ETH & XRP & LTC & BTC & ETH & XRP & LTC \\
        \hline
        BTC & 1.0000	&  &  &  & 1.0000 &  &  & \\
        ETH & 0.8467 & 1.0000 &  &  & 0.8416 & 1.0000 &  &  \\
        XRP & 0.6572 & 0.6796 & 1.0000 & & 0.6047 & 0.5772 & 1.0000 & \\
        LTC & 0.7801 & 0.8024 & 0.6979 & 1.0000 & 0.7644 & 0.8117 & 0.6511 & 1.0000 \\
        \hline
    \end{tabularx}
\end{table}

\subsection{Optimization process}

\begin{figure}[htbp]
    \refstepcounter{figure}
    \centering
    \includegraphics[width=\textwidth]{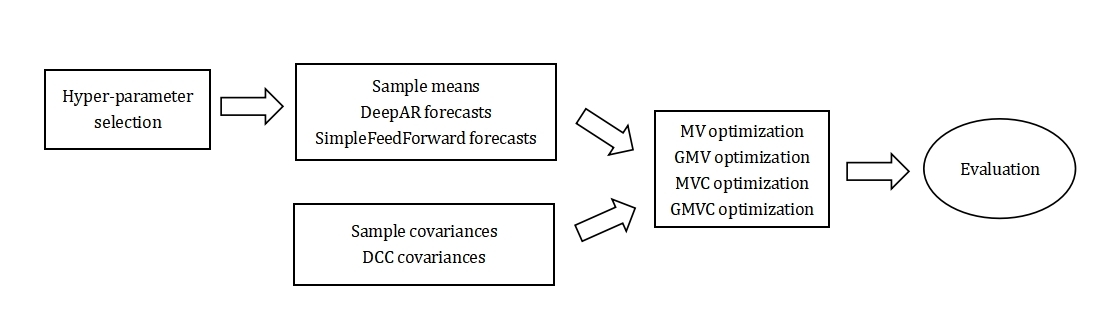}
    \caption*{Figure 1: Overall pipeline for portfolio optimization and performance evaluation.}
    \phantomsection\label{Figure 1}
\end{figure}

Now, we summarize the entire pipeline for portfolio selection as shown in \hyperref[Figure 1]{Figure 1}. First, we use the Optuna library to conduct hyper-parameter optimization for model training and forecasting, employing the aggregate RMSE as the optimization objective. Next, the model is trained using the selected optimal hyper-parameters to generate forecasts. At the same time, the sample means, sample covariances, and DCC covariances are estimated. Given these conditional mean and covariance estimates, the optimization frameworks described above are solved to obtain the optimal portfolio weights. Finally, the optimal weights are applied to the real-world dataset to evaluate the performance of the different portfolio strategies.

\subsection{Evaluation}
\subsubsection{Performance fee}
After the optimization process, we introduce a measure to evaluate the out-of-sample performance of the implemented strategies. A widely used performance measure in mean–variance analysis is the Sharpe ratio, a commonly used risk-adjusted indicator. However, according to \citet{della2009economic}, several studies argue that the Sharpe ratio severely underestimates the performance of dynamic asset allocation strategies (\citet{marquering2004economic} and \citet{han2006asset}), due to the overestimated conditional risk. Consequently, the evaluation criterion we adopt is the performance fee, defined as the additional fee applied to one of two strategies such that their average realized utilities become equal (\citep{fleming2001economic}). In this study, we compute the performance fees of various optimal portfolios relative to the naive 1/N rule. Following \citet{west1993utility}, we assume a quadratic utility which justifies the mean–variance analysis with a nonnormal return distribution, consistent with our data. Accordingly, following \citet{fleming2001economic}, the performance fee $\Phi$ is computed by solving the following equation:
\begin{align}
\sum_{t=1}^{Q}{ \{ (R_{p,t}^{*}-\Phi)  - \frac{\gamma}{2(1+\gamma)}(R_{p,t}^{*}-\Phi)^2 \} } = \sum_{t=1}^{Q}{ \{ R_{p,t}  - \frac{\gamma}{2(1+\gamma)}R_{p,t}^2 \} },
\label{pf1}
\end{align}
where $R_{p,t}^{*}=1+r_{p,t}$ with $r_{p,t}$ denoting the realized portfolio return net of transaction costs generated by the optimal portfolio strategy, and $R_{p,t}=1+r_{p,t}^{n}$ with $r_{p,t}^{n}$ denoting the realized portfolio return net of transaction costs obtained under the 1/N rule, which is independent of risk aversion. The parameter $\gamma$ represents the degree of relative risk aversion and is set equal to the value used in deriving $R_{p,t}^{*}$. The left side of Equation \ref{pf1} represents the utility achieved by the optimal strategy subject to a given performance fee, and the right side represents the utility achieved by the benchmark strategy. The performance fee has an intuitive economic interpretation: it indicates the maximum fee an investor is willing to pay to switch from the $1/N$ benchmark to the corresponding optimal portfolio strategy. Rearranging Equation \ref{pf1} by subtracting the right-hand side from the left-hand side yields:
\begin{align}
    -\Phi^2 + 2( \frac{\sum_{t=1}^{Q}{R_{p,t}^{*}}}{Q} - \frac{1+\gamma}{\gamma} )\Phi + 2\frac{1+\gamma}{\gamma}\frac{\sum_{t=1}^{Q}{( R_{p,t}^{*} - R_{p,t} )}}{Q} + \frac{\sum_{t=1}^{Q}{[R_{p,t}^2 - (R_{p,t}^{*})^2 ]}}{Q} = 0,
\label{pf2}
\end{align}
The left-hand side of Equation \ref{pf2} represents the amount by which the utility achieved by the optimal strategy subject to the performance fee exceeds that of the benchmark strategy, denoted by $\Delta U$. Solving this quadratic equation gives:
\begin{align}
    \Phi = \frac{B \pm \sqrt{B^2+4C}}{2}, 
\label{Phi}
\end{align}
where
\begin{align}
B &= 2( \frac{\sum_{t=1}^{Q}{R_{p,t}^{*}}}{Q} - \frac{1+\gamma}{\gamma} ), \\
C &= 2\frac{1+\gamma}{\gamma}\frac{\sum_{t=1}^{Q}{( R_{p,t}^{*} - R_{p,t} )}}{Q} + \frac{\sum_{t=1}^{Q}{[R_{p,t}^2 - (R_{p,t}^{*})^2 ]}}{Q}.
\label{BC}
\end{align}
As shown in \hyperref[Figure 2]{Figure 2}, $\Delta U$ is greater than 0 in the interval $(\Phi_1,\Phi_2)$, where $ \Phi_1 = \frac{B - \sqrt{B^2+4C}}{2} $ and $ \Phi_2 = \frac{B + \sqrt{B^2+4C}}{2} $ are the two solutions of the quadratic function. This means that an investor would be willing to pay a higher performance fee than $\Phi_1$ to switch from the 1/N benchmark to the corresponding optimal portfolio strategy, which contradicts the economic interpretation of the performance fee. Therefore, following \citet{kirby2012s}, we discard $ \Phi_1 $ and adopt $\Phi_2$ as the valid solution.
\par
\begin{figure}[htbp]
    \refstepcounter{figure}
    \centering
    \begin{tikzpicture}
    \begin{axis}[
        xlabel={$\Phi$},
        ylabel={$\Delta U$},
        axis lines=middle,
        enlarge x limits,
        enlarge y limits,
        width=7cm,
        height=5cm,
        xtick=\empty,
        ytick=\empty,
        every axis x label/.style={
            at={(ticklabel* cs:1)},
            anchor=west,
        },
        every axis y label/.style={
            at={(ticklabel* cs:1)},
            anchor=south,
        },
        clip=false
        ]
        \addplot[domain=-3:2, samples=100, color=blue]{-(x+0.7)^2+2.25};
        
        \node at (axis cs:-2.2,0) [anchor=south east] {$(\Phi_1, 0)$};
        \node at (axis cs:0.8,0) [anchor=south west] {$(\Phi_2, 0)$};
    \end{axis}
    \end{tikzpicture}
    \caption*{Figure 2: The excess utility ($\Delta U$) of the optimal strategy subject to a performance fee relative to the benchmark strategy.}
    \phantomsection\label{Figure 2}
\end{figure}
Further observing Equation \ref{Phi} $\sim$ \ref{BC}, a positive performance fee arises when $ B >0 $ and $ B^2 +4C \geq 0 $. However, this is rarely the case, because $B$ is typically negative. Thus, the conditions $\frac{\sum_{t=1}^{Q}{ R_{p,t}^{*} }}{Q} > \frac{\sum_{t=1}^{Q}{ R_{p,t} }}{Q} $ and $ \frac{\sum_{t=1}^{Q}{ (R_{p,t}^{*})^2 }}{Q} < \frac{\sum_{t=1}^{Q}{ R_{p,t}^2 }}{Q} $, which imply higher returns and lower risk relative to the benchmark, facilitate a positive performance fee. Rearranging Equation \ref{pf2} gives:
\begin{align}
-[\Phi - (\frac{\sum_{t=1}^{Q}{R_{p,t}^{*}}}{Q} - \frac{1+\gamma}{\gamma}) ]^2  + D + \frac{ \sum_{v \neq w}{2R_{p,v}^{*}R_{p,w}^{*}} }{Q^2} - \frac{(Q-1)\sum_{t=1}^{Q}{(R_{p,t}^{*})^2 } }{Q^2} = 0
\label{pf3}
\end{align}
where
\begin{align}
    D = (\frac{1+\gamma}{\gamma})^2 - 2\frac{1+\gamma}{\gamma}\frac{\sum_{t=1}^{Q}{R_{p,t}}}{Q} + \frac{\sum_{t=1}^{Q}{R_{p,t}^2}}{Q}
\end{align}
As shown in Equation \ref{pf3}, given the performance of the benchmark strategy, despite the existence of an uncertainty term ($\sum_{v \neq w}{2R_{p,v}^{*}R_{p,w}^{*}}$), however, an increase in return ($\sum_{t=1}^{Q}{R_{p,t}^{*}}/Q$) shifts the quadratic curve to the right, thereby increasing the performance fee. Similarly, a decrease in risk ($\sum_{t=1}^{Q}{(R_{p,t}^{*})^2}/Q$) shifts the curve upward, also improving the performance fee. Moreover, because a rightward shift has a more direct effect on determining the performance fee than an upward shift, return plays a more significant role than risk.

\subsubsection{Other metrics}
We evaluate the performance of different portfolio strategies over the entire out-of-sample period. As we derive optimal portfolios for risk-aversion coefficients $\gamma$ ranging from 1 to 10, we average the corresponding out-of-sample performance measures across different values of $\gamma$. We then define the average annualized performance fee $\overline{PF}$ as follows:
\begin{align}
     \overline{PF}=\frac{1}{10}\sum_{\gamma=1}^{10}{\kappa\Phi_2^{\gamma}},
\end{align}
where $\kappa$ denotes a multiplier that annualizes the performance fee and $\Phi_2^{\gamma}$ represents the adopted performance fee solution corresponding to a given value of $\gamma$.
\par

In addition to the performance fee, we also employ additional out-of-sample metrics to analyze the empirical results. As the out-of-sample length is the same for all strategies, we directly compare the total out-of-sample portfolio returns net of transaction costs. Similarly, averaging across values of $\gamma$ from 1 to 10, the average total out-of-sample portfolio return net of transaction costs $\overline{R_p}$ is defined as follows: 
\begin{align}
\overline{R_p} = \frac{1}{10}\sum_{\gamma=1}^{10}
\sum_{t=1}^{Q}{R_{p,t}^{\gamma}}, 
\end{align}
where $R_{p,t}^{\gamma}$ represents the portfolio return net of transaction costs obtained using a particular value of $\gamma$ at time t. 
\par

We also examine the risk profile of the portfolio and calculate the average total out-of-sample portfolio risk $\overline{R_p^2}$ as follows:
\begin{align}
\overline{R_p^2} = \frac{1}{10}\sum_{\gamma=1}^{10}
\sum_{t=1}^{Q}{(R_{p,t}^{\gamma})^2}, 
\end{align}
To further enrich the empirical analysis, we decompose the portfolio return net of transaction costs $\overline{R_p}$ into two components: the average total out-of-sample gross portfolio return $\overline{R_g}$ and the average total out-of-sample transaction costs $\overline{TC}$, and which can be expressed as:
\begin{align}
\overline{R_g} = \frac{1}{10}\sum_{\gamma=1}^{10}
\sum_{t=1}^{Q}{R_{g,t}^{\gamma}},\\
\overline{TC} = \frac{1}{10}\sum_{\gamma=1}^{10}\sum_{t=1}^{Q}{TC_t^{\gamma}},
\end{align}
where $R_{g,t}^{\gamma}$ and $TC_t^{\gamma}$ represent the gross portfolio return and the transaction cost obtained using a particular value of $\gamma$ at time t, respectively.
\par
The forecasting accuracy is evaluated using the aggregate RMSE measure defined above, replacing the validation set with the test set. Furthermore, we examine the variation characteristics of different moment estimators. The $L_2$-norm and the Frobenius norm are employed to measure the variation levels ($VL$) of the return forecasting series and the estimated covariance matrices, respectively, as follows:
\begin{align}
VL_{mean} = \frac{\sum_{t=1}^{Q-1}{|| \hat{r}_{t+1}-\hat{r}_{t} ||_2}}{Q-1} = \frac{\sum_{t=1}^{Q-1}{\sqrt{\sum_{i=1}^{N}{(\hat{r}_{i,t+1}-\hat{r}_{i,t})^2}}}}{Q-1}, \\
VL_{cov} = \frac{\sum_{t=1}^{Q-1}{|| \hat{\Sigma}_{t+1}-\hat{\Sigma}_{t} ||_F}}{Q-1} = \frac{\sum_{t=1}^{Q-1}{\sqrt{\sum_{i,j}{(\hat{\Sigma}_{ij,t+1}-\hat{\Sigma}_{ij,t})^2}}}}{Q-1}.
\end{align}

\begin{table}[htbp]
    \refstepcounter{table} 
    \renewcommand{\arraystretch}{1} 
    \caption*{Table 4: One-step-ahead forecasting accuracy and variation levels.}
    \phantomsection\label{Table 4}
    \begin{tabularx}{\textwidth}{c|XXXXX}
        \hline
        \multicolumn{6}{l}{Panel A: Forecasting accuracy.} \\
        \hline
            & SM & DA & SFF & sample cov & DCC cov \\
        \hline
        daily case & 0.0477 & 0.0479 & 0.0480 & -- & -- \\
        weekly case & 0.1252 & 0.1296 & 0.1292 & -- & -- \\    
    \end{tabularx}
    \begin{tabularx}{\textwidth}{c|XXXXX}
        \hline
        \multicolumn{6}{l}{Panel B: Variation levels.} \\
        \hline
            & SM & DA & SFF & sample cov & DCC cov \\
        \hline
        daily case & 0.0006 & 0.0162 & 0.0148 & 0.0001 & 0.0054 \\ 

        weekly case & 0.0113 & 0.1152 & 0.0088 & 0.0038 & 0.0370 \\ 
        \hline
    \end{tabularx}
    \begin{tabular}{p{\textwidth}}
    Note: ``SM'', ``DA'', and ``SFF'' indicate the sample means, DeepAR forecasts, and SimpleFeedForward forecasts, respectively.
    \end{tabular}
\end{table}

\section{Empirical results and discussion}

This section reports and discusses the empirical results for daily and weekly rebalancing cases. On a risk-adjusted basis, portfolio performance is determined by risk and return. We further decompose the return net of transaction costs $\overline{R_p}$ into gross return $\overline{R_g}$ and transaction costs $\overline{TC}$. Accordingly, we systematically analyze how different frameworks affect the three aspects (gross return $\overline{R_g}$, transaction costs $\overline{TC}$ and portfolio risk $\overline{R_p^2}$) that determine portfolio performance. 
\par

In detail, because the conditional mean estimates are associated with the term $ w_t'\mu_{t+1|t}$, we believe that accurate return forecasts should lead to the desired gross return measure ($\overline{R_g}$). Similarly, given that the conditional covariance estimates are associated with the term $w_t'\Sigma_{t+1|t}w_t$, it is reasonable to expect that accurate volatility forecasts should reduce portfolio risk ($\overline{R_p^2}$). In addition, greater instability in the estimators can amplify fluctuations in portfolio returns, which in turn increases portfolio risk. Finally, with respect to the transaction costs $\overline{TC}$, the turnover penalty term $ \beta||w_t - w_{t-1^{+}}^*||_{1} $ can inhibit excessive rebalancing and more volatile estimators also incur considerable transaction fees.\footnotemark[3]
\footnotetext[3]{In effect, the turnover penalty may also affect the gross return $\overline{R_g}$ and the risk $\overline{R_p^2}$ through discouraging excessive rebalancing. However, to avoid an overly detailed discussion, and because this is not the main focus of our study, we do not discuss these effects further here.}

\begin{table}[htbp]
    \refstepcounter{table}
    \renewcommand{\arraystretch}{1} 
    \caption*{Table 5: Average total out-of-sample portfolio returns net of transaction costs $\overline{R_p}$, average total out-of-sample portfolio risk $\overline{R_p^2}$, average total out-of-sample gross portfolio returns $\overline{R_g}$ and average total out-of-sample transaction costs $\overline{TC}$ achieved by the MV optimization framework, under daily and weekly rebalancing.}
    \phantomsection\label{Table 5}
    \begin{tabularx}{\textwidth}{c|X|XXX}
        \hline
        \multicolumn{5}{l}{Panel A: Average total out-of-sample portfolio returns net of transaction costs $\overline{R_p}$.} \\
        \hline
        & 1/N & SM & DA & SFF \\
        \hline
        daily & -0.2004 & -0.8756 & -5.0248 & -4.9959 \\ 
        weekly & -0.3538 & -0.8419 & -1.9713 & -0.5182 \\ 
        \hline
    \end{tabularx}
    \begin{tabularx}{\textwidth}{c|X|XXX}
        \multicolumn{5}{l}{Panel B: Average total out-of-sample portfolio risk $\overline{R_p^2}$.} \\
        \hline
        & 1/N & SM & DA & SFF \\
        \hline
        daily & 1.5291 & 1.2912 & 1.9159 & 1.7309 \\ 
        weekly & 1.4165 & 1.3784 & 1.4663 & 1.3201 \\ 
        \hline
    \end{tabularx}
    \begin{tabularx}{\textwidth}{c|X|XXX}
        \multicolumn{5}{l}{Panel C: Average total out-of-sample gross portfolio returns $\overline{R_g}$.} \\
        \hline
        & 1/N & SM & DA & SFF \\
        \hline
        daily & -0.1354 & -0.6103 & 0.5019 & 0.0358 \\ 
        weekly & -0.3238 & -0.7056 & -1.1329 & -0.3073 \\ 
        \hline
    \end{tabularx}
    \begin{tabularx}{\textwidth}{c|X|XXX}
        \multicolumn{5}{l}{Panel D: Average total out-of-sample transaction costs $\overline{TC}$.} \\
        \hline
        & 1/N & SM & DA & SFF \\
        \hline
        daily & 0.0651 & 0.2653 & 5.5267 & 5.0316 \\ 
        weekly & 0.0300 & 0.1363 & 0.8384 & 0.2108 \\ 
        \hline
    \end{tabularx}
\end{table}

\subsection{The results of the MV optimization framework}

We begin by examining the characteristics of the return forecasts series obtained from the different methods. The first one is the forecasting accuracy evaluated by the aggregate RMSE measure. As shown in Panel A of \hyperref[Table 4]{Table 4}, the forecasting accuracy of all methods is similar, suggesting that it is not the primary cause of the performance differences in optimal portfolios. These forecasting accuracy values (aggregate RMSE) correspond to a daily prediction error of nearly 5\% and a weekly prediction error of 12\% $\sim$ 13\%, which suggest considerable errors and that the mean predictive models considered have essentially no predictive ability. Accordingly, although the gross returns $\overline{R_g}$ vary across optimal strategies as shown in Panel C of \hyperref[Table 5]{Table 5}, we understand this as an outcome of specific data. Hence, we find that time-varying mean estimators offer no improvement over sample estimators in terms of asset allocation. 
\par

\begin{table}[htbp]
    \refstepcounter{table}
    \renewcommand{\arraystretch}{1} 
    \caption*{Table 6: Average annualized performance fees $\overline{PF}$ achieved by the MV optimization framework under daily and weekly rebalancing.}
    \phantomsection\label{Table 6}
    \begin{tabularx}{\textwidth}{c|XXX|XXX}
        \hline
        & \multicolumn{3}{l|}{Panel A: daily rebalancing} & \multicolumn{3}{l}{Panel B: weekly rebalancing} \\
        \cline{2-7}
            & SM & DA & SFF & SM & DA & SFF \\
        \hline
        $\overline{PF}$ & -0.2387 & -2.0798 & -2.0319 & -0.1931 & -0.6749 & -0.0459   \\ 
        \hline
    \end{tabularx}
    \begin{tabular}{p{\textwidth}}
    Note: ``SM'', ``DA'', and ``SFF'' indicate the sample means, DeepAR forecasts, and SimpleFeedForward forecasts, respectively.
    \end{tabular}
\end{table}

Subsequently, we further examine the variation in the characteristics of different return forecasting series. Panel B of \hyperref[Table 4]{Table 4} shows that the variation levels of the deep-learning forecasting series are substantially higher than those of the sample mean series. This is consistent with the observation in \citet{gao2018commodities} that time-varying moment forecasts exhibit obviously greater volatility than those obtained from simpler methods. One exception is the weekly SimpleFeedForward forecasts, which calculate the average of 1000 sample paths, and the SimpleFeedForward model, which is a relatively simple deep-learning model. A high level of estimator variation not only leads to more frequent rebalancing and consequently higher transaction costs, (see Panel D of \hyperref[Table 5]{Table 5}), but also increases portfolio risk, which can be confirmed in Panel B (with the same exception for weekly SimpleFeedForward forecasts).
\par

However, we calculate the actual variation levels of the test set data to be 0.1067 for daily returns and 0.2956 for weekly returns, which are even substantially higher than those of the deep-learning forecasts. That is, although the variation levels of the deep-learning forecasts are relatively close to the actual ones, their limited predictive ability for conditional means leads to frequent rebalancing that generates no additional gains in asset allocation but results in substantial transaction costs and higher portfolio risk. Then, we can draw a preliminary conclusion that time-varying mean estimators are inferior to sample estimators in portfolio optimization.
\par

Within the MV framework, we do not compare among covariance estimators. Therefore, we proceed to examine the performance of the optimal portfolio strategies relative to the naive $1/N$ benchmark. Our results show that all MV optimal strategies underperform naive diversification in terms of the performance fee, with some even yielding unreasonable results (see \hyperref[Table 6]{Table 6}). We observe that although some optimal strategies achieve higher gross returns ($R_g$) than the naive strategy, the associated transaction costs offset these advantages, resulting in all optimal strategies realizing lower net returns ($R_p$) than the naive benchmark (see Panel A and Panel C of \hyperref[Table 5]{Table 5}). Ultimately, as discussed earlier, the realized performance fees ($\overline{PF}$) primarily follow $\overline{R_p}$ rather than portfolio risk $\overline{R_p^2}$. As for those anomalous results, they are mainly caused by excessive transaction costs induced by highly volatile time-varying moment estimators.
\par

Our results show that the benefits of incorporating mean estimators into portfolio optimization are typically insufficient to offset their associated drawbacks. Therefore, we next consider the global minimum-variance strategy that excludes mean estimators. 
\par

\begin{table}[htbp]
    \refstepcounter{table}
    \renewcommand{\arraystretch}{1} 
    \caption*{Table 7: Average total out-of-sample portfolio returns net of transaction costs $\overline{R_p}$, average total out-of-sample portfolio risk $\overline{R_p^2}$, average total out-of-sample gross portfolio returns $\overline{R_g}$ and average total out-of-sample transaction costs $\overline{TC}$ achieved by the GMV optimization framework, under daily and weekly rebalancing.}
    \phantomsection\label{Table 7}
    \begin{tabularx}{\textwidth}{c|X|XX}
        \hline
        \multicolumn{4}{l}{Panel A: Average total out-of-sample portfolio returns net of transaction costs $\overline{R_p}$.} \\
        \hline
        & 1/N & sample cov & DCC cov \\
        \hline
        daily & -0.2004 & -0.2163 & -0.7593  \\ 
        weekly & -0.3538 & -0.1389 & -0.2706  \\ 
        \hline
    \end{tabularx}
    \begin{tabularx}{\textwidth}{c|X|XX}
        \multicolumn{4}{l}{Panel B: Average total out-of-sample portfolio risk $\overline{R_p^2}$.} \\
        \hline
        & 1/N & sample cov & DCC cov \\
        \hline
        daily & 1.5291 & 1.0512 & 1.0930 \\ 
        weekly & 1.4165 & 1.2205 & 1.2240 \\ 
        \hline
    \end{tabularx}
    \begin{tabularx}{\textwidth}{c|X|XX}
        \multicolumn{4}{l}{Panel C: Average total out-of-sample gross portfolio returns $\overline{R_g}$.} \\
        \hline
        & 1/N & sample cov & DCC cov \\
        \hline
        daily & -0.1354 & -0.1730 & -0.1344  \\ 
        weekly & -0.3238 & -0.0602 & -0.1174  \\ 
        \hline
    \end{tabularx}
    \begin{tabularx}{\textwidth}{c|X|XX}
        \multicolumn{4}{l}{Panel D: Average total out-of-sample transaction costs $\overline{TC}$.} \\
        \hline
        & 1/N & sample cov & DCC cov \\
        \hline
        daily & 0.0651 & 0.0433 & 0.6249 \\ 
        weekly & 0.0300 & 0.0787 & 0.1533 \\ 
        \hline
    \end{tabularx}
\end{table}

\subsection{The results of the GMV optimization framework}

We next discuss the empirical results of the GMV implementation. First, we compare the time-varying DCC covariance estimators with the sample covariance estimators. We believe that covariance estimators are primarily related to risk minimization in portfolio construction, while the differences in gross returns $\overline{R_g}$ across the GMV optimal strategies are likely due to data-specific randomness. 
\par

As shown in Panel B and Panel D of \hyperref[Table 7]{Table 7}, the GMV strategies employing DCC covariances fail to improve portfolio risk relative to those using sample covariances, regardless of daily or weekly rebalancing, yet they incur higher transaction costs. Therefore, similar to the mean estimators, the time-varying covariance estimators are also inferior to their sample counterparts. At this point, we can draw a preliminary conclusion that in portfolio optimization, time-varying moment estimators are inferior to sample estimators, both in mean and covariance estimation.
\par

As volatility timing strategies do not involve mean estimation, we next move to \hyperref[Table 8]{Table 8} to examine the performance of the GMV strategies relative to that of the naive $1/N$ benchmark. Our results show that most of the GMV strategies yield positive performance fees (except for the daily rebalanced strategy using DCC covariances), which aligns closely to previous studies such as \citet{demiguel2009optimal} and \citet{olivares2018robust}. The only exception can be attributed to the substantial transaction costs arising from the high variation level of the DCC covariances and frequent rebalancing. A closer inspection reveals that all GMV optimal strategies achieve lower portfolio risk than the naive $1/N$ benchmark, which helps explain their generally superior performance.
\par

Compared with the performance under the MV framework, our results further reinforce the view that excluding mean estimators from portfolio optimization is beneficial, and demonstrate the substantial economic value of predictability in return volatility in portfolio construction.
\par

\begin{table}[htbp]
    \refstepcounter{table} 
    \renewcommand{\arraystretch}{1} 
    \caption*{Table 8: Average annualized performance fees $\overline{PF}$ achieved by the GMV optimization framework under daily and weekly rebalancing.}
    \phantomsection\label{Table 8}
    \begin{tabularx}{\textwidth}{c|XX|XX}
        \hline
        & \multicolumn{2}{l|}{Panel A: daily rebalancing} & \multicolumn{2}{l}{Panel B: weekly rebalancing} \\
        \cline{2-5}
            & sample cov & DCC cov & sample cov & DCC cov \\
        \hline
        $\overline{PF}$ & \underline{0.0733} & -0.1614 & \underline{0.1217} & \underline{0.0664}  \\ 
        \hline
    \end{tabularx}
\end{table}

\begin{table}[htbp]
    \refstepcounter{table}
    \renewcommand{\arraystretch}{1} 
    \caption*{Table 9: Average total out-of-sample portfolio returns net of transaction costs $\overline{R_p}$, average total out-of-sample portfolio risk $\overline{R_p^2}$, average total out-of-sample gross portfolio returns $\overline{R_g}$ and average total out-of-sample transaction costs $\overline{TC}$ achieved by the MVC and GMVC optimization frameworks, under daily and weekly rebalancing.}
    \phantomsection\label{Table 9}
    \begin{tabularx}{\textwidth}{c|X|XXX|XX}
        \hline
        \multicolumn{7}{l}{Panel A: Average total out-of-sample portfolio returns net of transaction costs $\overline{R_p}$.} \\
        \hline
        & 1/N & SM & DA & SFF & sample cov & DCC cov \\
        \hline
        daily & -0.2004 & -0.4812 & -1.9421 & -0.3437 & -0.4607 & -0.0024 \\ 
        weekly & -0.3538 & -0.6632 & -1.5216 & -0.4742 & -0.3511 & -0.2157 \\ 
        \hline
    \end{tabularx}
    \begin{tabularx}{\textwidth}{c|X|XXX|XX}
        \multicolumn{7}{l}{Panel B: Average total out-of-sample portfolio risk $\overline{R_p^2}$.} \\
        \hline
        & 1/N & SM & DA & SFF & sample cov & DCC cov \\
        \hline
        daily & 1.5291 & 1.0292 & 1.4435 & 1.2213 & 1.0616 & 1.2061 \\ 
        weekly & 1.4165 & 1.3357 & 1.5649 & 1.2299 & 1.1484 & 1.2698 \\ 
        \hline
    \end{tabularx}
    \begin{tabularx}{\textwidth}{c|X|XXX|XX}
        \multicolumn{7}{l}{Panel C: Average total out-of-sample gross portfolio returns $\overline{R_g}$.} \\
        \hline
        & 1/N & SM & DA & SFF & sample cov & DCC cov \\
        \hline
        daily & -0.1354 & -0.4761 & -0.6647 & -0.1785 & -0.4556 & 0.0254 \\ 
        weekly & -0.3238 & -0.6170 & -0.7937 & -0.4548 & -0.3367 & -0.1444 \\ 
        \hline
    \end{tabularx}
    \begin{tabularx}{\textwidth}{c|X|XXX|XX}
        \multicolumn{7}{l}{Panel D: Average total out-of-sample transaction costs $\overline{TC}$.} \\
        \hline
        & 1/N & SM & DA & SFF & sample cov & DCC cov \\
        \hline
        daily & 0.0651 & 0.0050 & 1.2774 & 0.1652 & 0.0051 & 0.0278 \\ 
        weekly & 0.0300 & 0.0463 & 0.7279 & 0.0194 & 0.0144 & 0.0714  \\ 
        \hline
    \end{tabularx}
\end{table}

\subsection{The results with turnover penalty}
We next examine the empirical results of the two previous optimization frameworks with a turnover penalty imposed, namely the MVC and GMVC optimization frameworks. We first turn to Panels B, C, and D of \hyperref[Table 9]{Table 9}. Our results show that deep-learning forecasts still fail to improve asset allocation ($R_g$) relative to sample means. Moreover, even with the incorporation of a turnover penalty, they continue to incur higher transaction costs (except for the most stable weekly SimpleFeedForward forecasts). Similarly, the time-varying DCC covariances do not lead to lower portfolio risks compared with the sample covariances and tend to incur slightly higher transaction costs. Overall, these results further support our earlier conclusion that time-varying moment estimators underperform their sample counterparts in portfolio construction, even after incorporating a turnover penalty.
\par

We next analyze how the turnover penalty influences portfolio performance. Comparing Panel D of \hyperref[Table 9]{Table 9} with those of \hyperref[Table 5]{Table 5} and \hyperref[Table 7]{Table 7}, reveals that the turnover penalty substantially reduces transaction costs $\overline{TC}$ across all optimal portfolios, particularly for strategies that are frequently rebalanced, employing time-varying or mean estimators. To avoid an overly detailed analysis, and also because it is not the main focus of our study, we skip the discussion of how the turnover penalty affects gross returns $R_g$ and portfolio risks $\overline{R_p^2}$, and instead focus directly on its impact on performance fees. 
\par

Comparing \hyperref[Table 10]{Table 10} with \hyperref[Table 6]{Table 6} and \hyperref[Table 8]{Table 8}, we find that imposing a turnover penalty substantially improves portfolio performance, in terms of performance fees, for all MV optimal portfolios. This finding is consistent with \citet{yoshimoto1996mean} and \citet{olivares2018robust}, and underscores the importance of the turnover penalty in the mean–variance framework. However, in volatility timing strategies, the turnover penalty only improves the performance of strategies employing DCC covariances, but not those using sample covariances. This effectively corresponds to the results of a short-sale-constrained minimum-variance portfolio with nominal transaction costs in \citet{olivares2018robust}. 
\par

Furthermore, our empirical results show that the improvement brought by the turnover penalty diminishes as the rebalancing frequency declines. This observation aligns with our earlier analytical argument that a lower rebalancing frequency reduces the effect of imposing a turnover penalty.
\par

Likewise, we then investigate which optimal strategies are able to outperform the $1/N$ benchmark. The results in \hyperref[Table 10]{Table 10} indicate that almost all strategies employing mean estimates continue to underperform both the $1/N$ benchmark and the volatility-timing strategies, although a turnover penalty is imposed, confirming that the potential predictability of asset returns is difficult to exploit in practice. However, even in the highly correlated cryptocurrencies market, volatility-timing strategies generally outperform the naive $1/N$ benchmark, which reaffirms the economic value of the predictability in return volatility. Better performance may be achieved in more idiosyncratic markets or with more diversified portfolios. 
\par

The improvement over the naive $1/N$ strategy remains limited, partly due to the high correlations observed in the cryptocurrencies market. This also suggests that the gains from optimal allocation are constrained when relying solely on daily or weekly historical data.
\par

\begin{table}[htbp]
    \refstepcounter{table}
    \renewcommand{\arraystretch}{1} 
    \caption*{Table 10: Average annualized performance fees $\overline{PF}$ achieved by the MVC and GMVC optimization frameworks, under daily and weekly rebalancing.}
    \phantomsection\label{Table 10}
    \begin{tabularx}{\textwidth}{c|XXX|XX}
        \hline
        & SM & DA & SFF & sample cov & DCC cov \\
        \hline
        daily & -0.0343 & -0.7114 & -0.0043 & -0.0309 & \underline{0.1395} \\ 
        weekly & -0.1114 & -0.5055 & -0.0143 & \underline{0.0456} & \underline{0.0829} \\ 
        \hline
    \end{tabularx}
    \begin{tabular}{p{\textwidth}}
    Note: ``SM'', ``DA'', and ``SFF'' indicate the sample means, DeepAR forecasts, and SimpleFeedForward forecasts, respectively.
    \end{tabular}
\end{table}

\section{Conclusion}

This study conducts a comprehensive investigation into portfolio optimization within the cryptocurrencies market, emphasizing the role of time-varying moments and transaction costs. By incorporating both analytical derivations and empirical evidence under the mean–variance analysis, we compare four optimization frameworks (MV, GMV, MVC, and GMVC), and employ different rebalancing frequencies and risk-aversion coefficients to ensure the robustness of our results.
\par

Our empirical findings reveal that time-varying estimators of the first and second moments, such as deep-learning forecasts and DCC covariances, are inferior to their sample-based counterparts. These estimators tend to be more volatile, which results in higher turnover and transaction costs without improving asset allocation efficiency or risk minimization. 
\par

Regarding the turnover penalty, our results show that it effectively reduces transaction costs across all optimal strategies but does not necessarily enhance the final performance of every strategy. Both our analytical derivation and empirical evidence further demonstrate that the impact of imposing a turnover penalty diminishes as the rebalancing frequency decreases, consistent with the theoretical deduction of \citet{woodside2013portfolio}.
\par

Even in the highly correlated cryptocurrencies market, most volatility-timing strategies earn positive performance fees and outperform the naive $1/N$ benchmark, underscoring the economic value of volatility predictability for portfolio construction. Nevertheless, portfolio strategies depending on mean forecasts underperform both the volatility-timing strategies and the naive 1/N benchmark, reconfirming that the potential predictability of asset returns is difficult to exploit in practice.
\par

Overall, this study highlights the importance of incorporating transaction costs and volatility predictability into portfolio optimization but cautions against relying excessively on mean estimators or time-varying estimators obtained from complex models. Future research may extend this analysis to broader asset classes and more heterogeneous markets, explore nonlinear transaction cost structures, or integrate hybrid learning architectures to enhance the robustness and economic interpretability of portfolio optimization models.

\bibliographystyle{abbrvnat}
\bibliography{references}

\begin{thebibliography}{41}
\providecommand{\natexlab}[1]{#1}
\providecommand{\url}[1]{\texttt{#1}}
\expandafter\ifx\csname urlstyle\endcsname\relax
  \providecommand{\doi}[1]{doi: #1}\else
  \providecommand{\doi}{doi: \begingroup \urlstyle{rm}\Url}\fi

\bibitem[Ackermann et~al.(2017)Ackermann, Pohl, and Schmedders]{ackermann2017optimal}
F.~Ackermann, W.~Pohl, and K.~Schmedders.
\newblock Optimal and naive diversification in currency markets.
\newblock \emph{Management Science}, 63\penalty0 (10):\penalty0 3347--3360, 2017.

\bibitem[Ahmed et~al.(2016)Ahmed, Liu, and Valente]{ahmed2016can}
S.~Ahmed, X.~Liu, and G.~Valente.
\newblock Can currency-based risk factors help forecast exchange rates?
\newblock \emph{International Journal of Forecasting}, 32\penalty0 (1):\penalty0 75--97, 2016.

\bibitem[Alexandrov et~al.(2019)Alexandrov, Benidis, Bohlke-Schneider, Flunkert, Gasthaus, Januschowski, Maddix, Rangapuram, Salinas, Schulz, et~al.]{alexandrov2019gluonts}
A.~Alexandrov, K.~Benidis, M.~Bohlke-Schneider, V.~Flunkert, J.~Gasthaus, T.~Januschowski, D.~C. Maddix, S.~Rangapuram, D.~Salinas, J.~Schulz, et~al.
\newblock Gluonts: Probabilistic time series models in python.
\newblock \emph{arXiv preprint arXiv:1906.05264}, 2019.

\bibitem[Bariviera and Merediz-Sol{\`a}(2021)]{bariviera2021we}
A.~F. Bariviera and I.~Merediz-Sol{\`a}.
\newblock Where do we stand in cryptocurrencies economic research? a survey based on hybrid analysis.
\newblock \emph{Journal of Economic Surveys}, 35\penalty0 (2):\penalty0 377--407, 2021.

\bibitem[Bergstra et~al.(2011)Bergstra, Bardenet, Bengio, and K{\'e}gl]{bergstra2011algorithms}
J.~Bergstra, R.~Bardenet, Y.~Bengio, and B.~K{\'e}gl.
\newblock Algorithms for hyper-parameter optimization.
\newblock \emph{Advances in neural information processing systems}, 24, 2011.

\bibitem[Cheikh et~al.(2020)Cheikh, Zaied, and Chevallier]{cheikh2020asymmetric}
N.~B. Cheikh, Y.~B. Zaied, and J.~Chevallier.
\newblock Asymmetric volatility in cryptocurrency markets: New evidence from smooth transition garch models.
\newblock \emph{Finance Research Letters}, 35:\penalty0 101293, 2020.

\bibitem[Chen et~al.(2021)Chen, Zhang, Mehlawat, and Jia]{chen2021mean}
W.~Chen, H.~Zhang, M.~K. Mehlawat, and L.~Jia.
\newblock Mean--variance portfolio optimization using machine learning-based stock price prediction.
\newblock \emph{Applied Soft Computing}, 100:\penalty0 106943, 2021.

\bibitem[Chopra and Ziemba(1993)]{chopra1993effect}
V.~K. Chopra and W.~T. Ziemba.
\newblock The effect of errors in means, variances, and covariances on optimal portfolio choice.
\newblock \emph{The Journal of Portfolio Management}, 19\penalty0 (2):\penalty0 6--11, 1993.

\bibitem[Della~Corte et~al.(2009)Della~Corte, Sarno, and Tsiakas]{della2009economic}
P.~Della~Corte, L.~Sarno, and I.~Tsiakas.
\newblock An economic evaluation of empirical exchange rate models.
\newblock \emph{The review of financial studies}, 22\penalty0 (9):\penalty0 3491--3530, 2009.

\bibitem[DeMiguel et~al.(2009)DeMiguel, Garlappi, and Uppal]{demiguel2009optimal}
V.~DeMiguel, L.~Garlappi, and R.~Uppal.
\newblock Optimal versus naive diversification: How inefficient is the 1/n portfolio strategy?
\newblock \emph{The review of Financial studies}, 22\penalty0 (5):\penalty0 1915--1953, 2009.

\bibitem[Du(2022)]{du2022mean}
J.~Du.
\newblock Mean--variance portfolio optimization with deep learning based-forecasts for cointegrated stocks.
\newblock \emph{Expert Systems with Applications}, 201:\penalty0 117005, 2022.

\bibitem[D’Hondt et~al.(2020)D’Hondt, De~Winne, Ghysels, and Raymond]{d2020artificial}
C.~D’Hondt, R.~De~Winne, E.~Ghysels, and S.~Raymond.
\newblock Artificial intelligence alter egos: Who might benefit from robo-investing?
\newblock \emph{Journal of Empirical Finance}, 59:\penalty0 278--299, 2020.

\bibitem[Engle(2002)]{engle2002dynamic}
R.~Engle.
\newblock Dynamic conditional correlation: A simple class of multivariate generalized autoregressive conditional heteroskedasticity models.
\newblock \emph{Journal of Business \& Economic Statistics}, 20\penalty0 (3):\penalty0 339--350, 2002.

\bibitem[Fleming et~al.(2001)Fleming, Kirby, and Ostdiek]{fleming2001economic}
J.~Fleming, C.~Kirby, and B.~Ostdiek.
\newblock The economic value of volatility timing.
\newblock \emph{The Journal of Finance}, 56\penalty0 (1):\penalty0 329--352, 2001.

\bibitem[Fleming et~al.(2003)Fleming, Kirby, and Ostdiek]{fleming2003economic}
J.~Fleming, C.~Kirby, and B.~Ostdiek.
\newblock The economic value of volatility timing using “realized” volatility.
\newblock \emph{Journal of Financial Economics}, 67\penalty0 (3):\penalty0 473--509, 2003.

\bibitem[Fousekis and Grigoriadis(2021)]{fousekis2021directional}
P.~Fousekis and V.~Grigoriadis.
\newblock Directional predictability between returns and volume in cryptocurrencies markets.
\newblock \emph{Studies in Economics and Finance}, 38\penalty0 (4):\penalty0 693--711, 2021.

\bibitem[Gao and Nardari(2018)]{gao2018commodities}
X.~Gao and F.~Nardari.
\newblock Do commodities add economic value in asset allocation? new evidence from time-varying moments.
\newblock \emph{Journal of Financial and Quantitative Analysis}, 53\penalty0 (1):\penalty0 365--393, 2018.

\bibitem[Golnari et~al.(2024)Golnari, Komeili, and Azizi]{golnari2024probabilistic}
A.~Golnari, M.~H. Komeili, and Z.~Azizi.
\newblock Probabilistic deep learning and transfer learning for robust cryptocurrency price prediction.
\newblock \emph{Expert Systems with Applications}, page 124404, 2024.

\bibitem[Han(2006)]{han2006asset}
Y.~Han.
\newblock Asset allocation with a high dimensional latent factor stochastic volatility model.
\newblock \emph{The Review of Financial Studies}, 19\penalty0 (1):\penalty0 237--271, 2006.

\bibitem[Hautsch and Voigt(2019)]{hautsch2019large}
N.~Hautsch and S.~Voigt.
\newblock Large-scale portfolio allocation under transaction costs and model uncertainty.
\newblock \emph{Journal of Econometrics}, 212\penalty0 (1):\penalty0 221--240, 2019.

\bibitem[Januschowski et~al.(2020)Januschowski, Gasthaus, Wang, Salinas, Flunkert, Bohlke-Schneider, and Callot]{januschowski2020criteria}
T.~Januschowski, J.~Gasthaus, Y.~Wang, D.~Salinas, V.~Flunkert, M.~Bohlke-Schneider, and L.~Callot.
\newblock Criteria for classifying forecasting methods.
\newblock \emph{International Journal of Forecasting}, 36\penalty0 (1):\penalty0 167--177, 2020.

\bibitem[Kan and Zhou(2007)]{kan2007optimal}
R.~Kan and G.~Zhou.
\newblock Optimal portfolio choice with parameter uncertainty.
\newblock \emph{Journal of Financial and Quantitative Analysis}, 42\penalty0 (3):\penalty0 621--656, 2007.

\bibitem[Katsiampa et~al.(2019)Katsiampa, Corbet, and Lucey]{katsiampa2019high}
P.~Katsiampa, S.~Corbet, and B.~Lucey.
\newblock High frequency volatility co-movements in cryptocurrency markets.
\newblock \emph{Journal of International Financial Markets, Institutions and Money}, 62:\penalty0 35--52, 2019.

\bibitem[Kirby and Ostdiek(2012)]{kirby2012s}
C.~Kirby and B.~Ostdiek.
\newblock It’s all in the timing: simple active portfolio strategies that outperform naive diversification.
\newblock \emph{Journal of financial and quantitative analysis}, 47\penalty0 (2):\penalty0 437--467, 2012.

\bibitem[Kraaijeveld and De~Smedt(2020)]{kraaijeveld2020predictive}
O.~Kraaijeveld and J.~De~Smedt.
\newblock The predictive power of public twitter sentiment for forecasting cryptocurrency prices.
\newblock \emph{Journal of International Financial Markets, Institutions and Money}, 65:\penalty0 101188, 2020.

\bibitem[Kynigakis and Panopoulou(2022)]{kynigakis2022does}
I.~Kynigakis and E.~Panopoulou.
\newblock Does model complexity add value to asset allocation? evidence from machine learning forecasting models.
\newblock \emph{Journal of Applied Econometrics}, 37\penalty0 (3):\penalty0 603--639, 2022.

\bibitem[Li et~al.(2024)Li, Chen, Zhou, Yang, and Zeng]{li2024deepar}
J.~Li, W.~Chen, Z.~Zhou, J.~Yang, and D.~Zeng.
\newblock Deepar-attention probabilistic prediction for stock price series.
\newblock \emph{Neural Computing and Applications}, pages 1--18, 2024.

\bibitem[Liu and Tsyvinski(2021)]{liu2021risks}
Y.~Liu and A.~Tsyvinski.
\newblock Risks and returns of cryptocurrency.
\newblock \emph{The Review of Financial Studies}, 34\penalty0 (6):\penalty0 2689--2727, 2021.

\bibitem[Ma et~al.(2021)Ma, Han, and Wang]{ma2021portfolio}
Y.~Ma, R.~Han, and W.~Wang.
\newblock Portfolio optimization with return prediction using deep learning and machine learning.
\newblock \emph{Expert Systems with Applications}, 165:\penalty0 113973, 2021.

\bibitem[Makridakis et~al.(2023)Makridakis, Spiliotis, Assimakopoulos, Semenoglou, Mulder, and Nikolopoulos]{makridakis2023statistical}
S.~Makridakis, E.~Spiliotis, V.~Assimakopoulos, A.-A. Semenoglou, G.~Mulder, and K.~Nikolopoulos.
\newblock Statistical, machine learning and deep learning forecasting methods: Comparisons and ways forward.
\newblock \emph{Journal of the Operational Research Society}, 74\penalty0 (3):\penalty0 840--859, 2023.

\bibitem[Marquering and Verbeek(2004)]{marquering2004economic}
W.~Marquering and M.~Verbeek.
\newblock The economic value of predicting stock index returns and volatility.
\newblock \emph{Journal of Financial and Quantitative Analysis}, 39\penalty0 (2):\penalty0 407--429, 2004.

\bibitem[Michaud(1989)]{michaud1989markowitz}
R.~O. Michaud.
\newblock The markowitz optimization enigma: Is ‘optimized’optimal?
\newblock \emph{Financial analysts journal}, 45\penalty0 (1):\penalty0 31--42, 1989.

\bibitem[Nelson(1991)]{nelson1991conditional}
D.~B. Nelson.
\newblock Conditional heteroskedasticity in asset returns: A new approach.
\newblock \emph{Econometrica: Journal of the econometric society}, pages 347--370, 1991.

\bibitem[Olivares-Nadal and DeMiguel(2018)]{olivares2018robust}
A.~V. Olivares-Nadal and V.~DeMiguel.
\newblock A robust perspective on transaction costs in portfolio optimization.
\newblock \emph{Operations Research}, 66\penalty0 (3):\penalty0 733--739, 2018.

\bibitem[Opie and Riddiough(2020)]{opie2020global}
W.~Opie and S.~J. Riddiough.
\newblock Global currency hedging with common risk factors.
\newblock \emph{Journal of Financial Economics}, 136\penalty0 (3):\penalty0 780--805, 2020.

\bibitem[Platanakis et~al.(2018)Platanakis, Sutcliffe, and Urquhart]{platanakis2018optimal}
E.~Platanakis, C.~Sutcliffe, and A.~Urquhart.
\newblock Optimal vs na{\"\i}ve diversification in cryptocurrencies.
\newblock \emph{Economics Letters}, 171:\penalty0 93--96, 2018.

\bibitem[Salinas et~al.(2020)Salinas, Flunkert, Gasthaus, and Januschowski]{salinas2020deepar}
D.~Salinas, V.~Flunkert, J.~Gasthaus, and T.~Januschowski.
\newblock Deepar: Probabilistic forecasting with autoregressive recurrent networks.
\newblock \emph{International Journal of Forecasting}, 36\penalty0 (3):\penalty0 1181--1191, 2020.

\bibitem[Tu and Zhou(2011)]{tu2011markowitz}
J.~Tu and G.~Zhou.
\newblock Markowitz meets talmud: A combination of sophisticated and naive diversification strategies.
\newblock \emph{Journal of Financial Economics}, 99\penalty0 (1):\penalty0 204--215, 2011.

\bibitem[West et~al.(1993)West, Edison, and Cho]{west1993utility}
K.~D. West, H.~J. Edison, and D.~Cho.
\newblock A utility-based comparison of some models of exchange rate volatility.
\newblock \emph{Journal of international economics}, 35\penalty0 (1-2):\penalty0 23--45, 1993.

\bibitem[Woodside-Oriakhi et~al.(2013)Woodside-Oriakhi, Lucas, and Beasley]{woodside2013portfolio}
M.~Woodside-Oriakhi, C.~Lucas, and J.~E. Beasley.
\newblock Portfolio rebalancing with an investment horizon and transaction costs.
\newblock \emph{Omega}, 41\penalty0 (2):\penalty0 406--420, 2013.

\bibitem[Yoshimoto(1996)]{yoshimoto1996mean}
A.~Yoshimoto.
\newblock The mean-variance approach to portfolio optimization subject to transaction costs.
\newblock \emph{Journal of the Operations Research Society of Japan}, 39\penalty0 (1):\penalty0 99--117, 1996.

\end{thebibliography}

\appendix

\section*{Appendix}

\subsection*{Appendix A: Proof of Proposition 1}

\begin{proof}
Let $w_t^{MVC,*}$ denote the solution to the MVC optimization problem, which must then satisfy the first-order conditions:
\begin{align}
\gamma\Sigma_{t|t+1} w_t^{MVC,*} + \beta g^* - \mu_{t|t+1} - \lambda l &= 0, \\
l'w_t^{MVC,*} - 1 &= 0,
\end{align}
where $ g^* $ is the subgradient vector of function $|| w_t-w_{t-1^+}^* ||_1$ evaluated at $w_t^*$ and $\lambda$ is the Lagrange multiplier. Solving for $w_t^{MVC,*}$ yields,
\begin{align}
w_t^{MVC,*} = \frac{1}{\gamma}(\Sigma_{t|t+1}^{-1} - \frac{\Sigma_{t|t+1}^{-1}ll'\Sigma_{t|t+1}^{-1}}{l'\Sigma_{t|t+1}^{-1}l} )(\mu_{t|t+1}-\beta g^*) + \frac{\Sigma_{t|t+1}^{-1}l}{l'\Sigma_{t|t+1}^{-1}l},
\end{align}
Let $\tilde{w}_t^*$ denote the solution of the MV optimization problem with the mean estimator $\tilde{\mu}_{t|t+1}=\mu_{t|t+1}-\beta g^*$, then using the classical efficient portfolio representation, we have:
\begin{align}
\tilde{w}_t^* &= \frac{1}{\gamma}(\Sigma_{t|t+1}^{-1} - \frac{\Sigma_{t|t+1}^{-1}ll'\Sigma_{t|t+1}^{-1}}{l'\Sigma_{t|t+1}^{-1}l} )\tilde{\mu}_{t|t+1} + \frac{\Sigma_{t|t+1}^{-1}l}{l'\Sigma_{t|t+1}^{-1}l} \\
    &= \frac{1}{\gamma}(\Sigma_{t|t+1}^{-1} - \frac{\Sigma_{t|t+1}^{-1}ll'\Sigma_{t|t+1}^{-1}}{l'\Sigma_{t|t+1}^{-1}l} )(\mu_{t|t+1}-\beta g^*) + \frac{\Sigma_{t|t+1}^{-1}l}{l'\Sigma_{t|t+1}^{-1}l}.
\end{align}
Consequently, $w_t^{MVC,*}=\tilde{w}_t^* $, and Proposition 1 holds.
\end{proof}

\subsection*{Appendix B: Implementation details for hyper-parameter selection}
For the DeepAR model, we consider seven hyper-parameters: ``num\_layers'' which represents the number of RNN layers; ``hidden\_size'' which represents the number of RNN cells for each layer; ``batch\_size'' which represents the size of the batches used for training; ``max\_epochs'' which is a part of the hyper-parameter ``trainer\_kwargs'' and represents an additional argument to provide to pl.Trainer for construction; ``num\_batches\_per\_epoch'' which represents the number of batches to be processed in each training epoch; ``lr'' which defines the learning rate; and ``num\_samples'' which is the number of samples to draw on the model. Six of these hyper-parameters are defined as categorical hyper-parameters, with their respective search spaces listed as $[1, 2, 3]$ for ``num\_layers'', $[2, 4, 8, 16]$ for ``hidden\_size'', $[2, 4, 8, 16]$ for ``batch\_size'', $[8, 16, 32, 64, 128]$ for ``max\_epochs'', $[2, 4, 8, 16]$ for ``num\_batches\_per\_epoch'', and $[10, 100, 1000]$ for ``num\_samples''. While the remaining hyper-parameter, ``lr'', is defined as floating point (log) hyper-parameter ranging from $5e^{-4}$ to $5e^{-3}$.
\par

For the SimpleFeedForward model, we consider six key hyper-parameters: ``batch\_size'', ``max\_epochs'', ``num\_batches\_per\_epoch'', ``lr'' and ``num\_samples'' which are similar to those in the DeepAR model; and ``hidden\_dimensions'' representing the size of hidden layers in the feed-forward network. Descriptions of these hyper-parameters are taken from \url{https://ts.gluon.ai/stable/index.html}. The search spaces are defined as follows: $[4, 8, 16, 32]$ for ``batch\_size'', $[16, 32, 64, 128, 256]$ for ``max\_epochs'', $[4, 8, 16, 32]$ for ``num\_batches\_per\_epoch'', and $[10, 100, 1000]$ for ``num\_samples''. The  floating point (log) hyper-parameter ``lr'' ranges between $5e^{-4}$ and $5e^{-3}$ and a series of indicative sizes are considered for ``hidden\_dimensions''. The selected hyper-parameter values are reported in \hyperref[Table A1]{Table A1} and \hyperref[Table A2]{Table A2}.

\begin{table}[htbp]
    \refstepcounter{table}
    \renewcommand{\arraystretch}{1} 
    \caption*{Table A1: Selected hyper-parameter values for the DeepAR model on daily and weekly data.}
    \phantomsection\label{Table A1}
    \begin{tabularx}{\textwidth}{XXX}
        \hline
        & daily & weekly \\
        \hline
        num\_layers & 1 & 1 \\
        hidden\_size & 8 & 16 \\
        batch\_size & 16 & 2 \\
        max\_epochs & 64 & 16 \\
        num\_batches\_per\_epoch & 4 & 2 \\
        lr & 0.0022 & 0.0036 \\
        num\_samples & 100 & 10 \\
        \hline
    \end{tabularx}
\end{table}

\begin{table}[htbp]
    \refstepcounter{table}
    \renewcommand{\arraystretch}{1} 
    \caption*{Table A2: Selected hyper-parameter values for the SimpleFeedForward model on daily and weekly data.}
    \phantomsection\label{Table A2}
    \begin{tabularx}{\textwidth}{XXX}
        \hline
        & daily & weekly \\
        \hline
        hidden\_dimensions & [16] & [2, 2] \\
        batch\_size & 8 & 32 \\
        max\_epochs & 128 & 32 \\
        num\_batches\_per\_epoch & 4 & 4 \\
        lr & 0.0037 & 0.0016 \\
        num\_samples & 100 & 1000 \\
        \hline
    \end{tabularx}
\end{table}

\end{document}